\documentclass[12pt]{article}

\ifx\pdfoutput\undefined
\usepackage[dvips,bookmarks]{hyperref}
\else
\usepackage{hyperref}
\fi
\hypersetup{colorlinks=false,bookmarksopen,bookmarksnumbered,citecolor=blue,
   pdfstartview=FitH}

\usepackage[dvips]{graphicx}
\usepackage{latexsym}

\usepackage{amssymb,amsfonts,amsmath}
\usepackage{graphicx} 
\usepackage{indentfirst}

 \usepackage{bbm}

\parskip=\medskipamount

\newcommand {\cD}{{\cal D}}

\newcommand {\cN}{{\cal N}}

\newcommand {\cQ}{{\cal Q}}

\newcommand {\cW}{{\cal W}}
\newcommand {\cX}{{\cal X}}

\def\a{\alpha}
\def \bi{\bibitem}

\def\b{\beta}

\def\d{\delta}
\def\e{\epsilon}

\def\g{\gamma}
\def\G{\Gamma}

\def\l{\lambda}

\def\o{\omega}

\def\q{\theta}

\def\s{\sigma}

\def\x{\xi}
\def\z{\zeta}
\def\D{\Delta}
\def\F{\Phi}

\def\L{\Lambda}
\def\O{\Omega}

\newcommand{\ad}{{\dot{\alpha}}}                           
\newcommand{\bd}{{\dot{\beta}}}                            
\newcommand{\ve}{\varepsilon}                            

\newcommand{\pa}{\partial}                           
\newcommand{\hf}{\frac12}

\newcommand{\vf}{\varphi}

\newcommand{\be}{\begin{equation}}
\newcommand{\ee}{\end{equation}}
\newcommand{\bea}{\begin{eqnarray}}
\newcommand{\eea}{\end{eqnarray}}
\newcommand{\non}{\nonumber}
\newcommand{\bm}[1]{\mbox{\boldmath$#1$}}

\def\double #1{#1{\hbox{\kern-2pt $#1$}}}

\begin{document}
\begin{titlepage}
\begin{flushright}
November, 2009 \\
\end{flushright}
\vspace{5mm}

\begin{center}
{\Large \bf  The Fayet-Iliopoulos term and nonlinear self-duality}
\\ 
\end{center}

\begin{center}

{\bf
Sergei M. Kuzenko\footnote{kuzenko@cyllene.uwa.edu.au}
} \\
\vspace{5mm}

\footnotesize{
{\it School of Physics M013, The University of Western Australia\\
35 Stirling Highway, Crawley W.A. 6009, Australia}}  
~\\
\vspace{2mm}

\end{center}
\vspace{5mm}

\begin{abstract}
\baselineskip=14pt
The  $\cN=1$ supersymmetric Born-Infeld action is known to describe the vector Goldstone multiplet 
for partially broken  $\cN=2$ rigid supersymmetry, and this model is believed to be unique.
However, it can be deformed by adding the 
Fayet-Iliopoulos term without losing the second nonlinearly realized supersymmetry. 
Although the first supersymmetry then becomes spontaneously broken, 
the deformed action still describes partial $\cN=2 \to \cN=1$ supersymmetry breaking.  
The unbroken supercharges in this theory correspond to a different choice of $\cN=1$ 
subspace in the $\cN=2$ superspace, as compared with the undeformed case.
Implications of the Fayet-Iliopoulos term for general models for self-dual nonlinear supersymmetric 
 electrodynamics are discussed.  The known ubiquitous appearance 
of the Volkov-Akulov action in such models is explained. We also present 
 a two-parameter  duality-covariant deformation of the $\cN=1$ supersymmetric Born-Infeld action
as a model for partial breaking of $\cN=2$ supersymmetry.

\end{abstract}
\vspace{1cm}

\vfill
\end{titlepage}

\newpage
\renewcommand{\thefootnote}{\arabic{footnote}}
\setcounter{footnote}{0}


Recently, there have been  interesting  discussions 
\cite{KS,DT} of the supercurrent \cite{FZ} (i.e. the multiplet of currents containing the 
energy-momentum tensor and the supersymmetry current)
in $\cN=1$ supersymmetric gauge theories with a Fayet-Iliopoulos (FI) term \cite{FI}. 
These works are 
primarily targeted at phenomenological applications of supergravity theories.
In the present note, we would like to elaborate on
somewhat different and  more formal aspects such as  implications of the FI term 
for partial supersymmetry breaking, and more generally in the context of  models 
for self-dual nonlinear supersymmetric electrodynamics.
In particular, we will revisit the conclusion of \cite{BG,RT} 
about uniqueness of the Goldstone-Maxwell multiplet model for partially 
broken $\cN=2$ supersymmetry.
At the end of this note, we will also comment on the claim made in \cite{DT} that 
``no supercurrent supermultiplet exists for globally supersymmetric gauge theories with non-zero 
Fayet-Iliopoulos terms.''

The  $\cN=1$ supersymmetric Born-Infeld (BI) action is\footnote{We follow the notation 
and conventions adopted in \cite{WB,BK}. The superspace integration measures in (\ref{sbi}) 
are defined as
follows: ${\rm d}^6z := {\rm d}^4 x \, {\rm d}^2 \q$,  ${\rm d}^6{\bar z} := {\rm d}^4 x \, {\rm d}^2 {\bar \q}$ 
and $ {\rm d}^8z := {\rm d}^4 x \, {\rm d}^2 \q \,{\rm d}^2 {\bar \q}$.
} 
\bea
S_{\rm SBI} [W]&=&
  \frac{1}{4}\int {\rm d}^6z \, W^2 +
\frac{1}{4}\int {\rm d}^6{\bar z} \,{\bar  W}^2
+  \frac{ g^2 }{ 4}   \int {\rm d}^8z \, \frac{W^2\,{\bar W}^2  }
{ 1 + \hf\, A \, +
\sqrt{1 + A +\frac{1}{4} \,B^2} }~,
\non  \\
  A &=&   \frac{ g^2 }{ 8} \,
\Big(D^2\,W^2 + {\bar D}^2\, {\bar W}^2 \Big)~,
\qquad
B =  \frac{ g^2 }{ 8} \, \Big(D^2\,W^2 -
{\bar D}^2\, {\bar W}^2 \Big)~. ~~~
\label{sbi}
\eea
Here $g$ is the coupling constant, 
$W_\a$  the chiral field strength of an Abelian vector multiplet, 
\bea
W_\a =-\frac{1}{4} {\bar D}^2 D_\a V~, \qquad V=\bar V~, 
\eea
with $V$ the gauge prepotential.
The action (\ref{sbi}) was  introduced for the first time in Refs.  \cite{DP,CF}
as a supersymmetric extension of the BI theory \cite{BornI},
and as such it is not unique.
Bagger and Galperin \cite{BG}, and  later Ro\v{c}ek and
Tseytlin \cite{RT}, using alternative techniques, discovered that 
$S_{\rm SBI} $ is the  action for a Goldstone-Maxwell multiplet
associated with  $\cN=2 \to \cN=1$ partial supersymmetry breaking.
This action was argued to be unique \cite{BG,RT}.
Being manifestly $\cN=1$ supersymmetric, $S_{\rm SBI} $  also proves to be invariant
under  a second, nonlinearly realized supersymmetry transformation
\bea
\d W_\a &=& \eta_\a
+ \frac{g^2}{4} \Big( 
 \frac{1}{4}\, {\bar D}^2 {\bar X}\, \eta_\a
+ {\rm i}\,  \pa_{\a \ad} X  \, {\bar \eta}^\ad \Big)~,
\label{2nd-var}
\eea
with $\eta_\a$ a constant spinor parameter.
Here $X$ is a chiral superfield, ${\bar D}_\ad X=0$,  satisfying
the nonlinear constraint \cite{BG,RT}
\be
X +\frac{g^2}{16} \,  X\, {\bar D}^2  
{\bar X}  = W^2~.
\label{n=1constraint}
\ee
The proof of the invariance is based on the observations \cite{BG,RT}  that 
(i) the functional (\ref{sbi}) can be rewritten in the form
\be
S_{\rm SBI} [W]= \frac{1}{4}\int {\rm d}^6z \, X +
\frac{1}{4}\int {\rm d}^6{\bar z} \,{\bar  X}~,
\label{bi-2}
\ee
and (ii) the chiral scalar $X$ transforms under (\ref{2nd-var}) as 
\bea
\d X &=& 2\eta^\a \, W_\a ~.
\label{1st-var} 
\eea

Consider now the  $\cN=1$ supersymmetric FI term \cite{FI}
\bea
S_{\rm FI} &=&
2\x \int {\rm d}^8z \,V 
=  \frac{\x}{2 }\int {\rm d}^6z \, \q^\a W_\a  +
\frac{\x}{2}\int {\rm d}^6{\bar z} \,{\bar \q}_\ad {\bar W}^\ad~.
\label{FI-twoforms}
\eea
It is easy to see that $S_{\rm FI} $ is also invariant under the second nonlinearly 
realized supersymmetry (\ref{2nd-var}), as pointed out recently in \cite{ADM}.\footnote{This property 
is directly related to the fact that that the $\cN=1$ FI term  preserves also $\cN=2$ 
supersymmetry \cite{Fayet}.}
Therefore, the theory with action 
\be
S= S_{\rm SBI} [W]+ S_{\rm FI} 
\label{S}
\ee
is manifestly $\cN=1$ supersymmetric, and is invariant under the second nonlinearly 
realized supersymmetry (\ref{2nd-var}).
We are going to show that the deformed action also describes partial $\cN=2 \to \cN=1$ 
supersymmetry breaking.

Let us study the bosonic sector of the component Lagrangian.
The component fields contained in $W_\a$ are:
\begin{subequations}
\bea
\l_\a (x) & = & W_\a |_{\q =0}~, \\
F_{\a \b} (x) &=& -\frac{\rm i}{4}
( D_\a W_\b + D_\b W_\a )|_{\q = 0}~,  \\
\cD(x) &=& -\hf D^\a W_\a |_{\q = 0}~,
\eea
\end{subequations}
with
\be
F_{\a \ad \, \b \bd } \equiv (\s^a)_{\a \ad}
(\s^b)_{\b \bd} F_{ab} = 2 \ve_{\a \b} \,
{\bar F}_{\ad \bd} + 2 \ve_{\ad \bd} \, F_{\a \b} ~
\ee
the electromagnetic field strength.
Setting the photino to zero, $\l_\a =0$,  the  bosonic Lagrangian can be shown to be 
\bea
L_{\rm boson}=   \frac{1}{g^2} \left\{ 
1 - \sqrt{1 + g^2 ({\bm u} + {\bar {\bm u}} )
+{1 \over 4}g^4 ({\bm u} - {\bar {\bm u}} )^2 } 
\right\} + \x \, \cD ~,
\label{L-bosonic}
\eea
where we have defined 
\bea
{\bm u} &:=& \frac{1}{8} D^2 W^2 |_{\q_\a = \l_\a=0} = \o - \hf \cD^2~, \qquad
\o :=  \frac{1}{4} \, F^{ab} F_{ab} + \frac{\rm i}{4} \, F^{ab} \tilde{F}_{ab} ~,
\label{bm-u}
\eea
with $ \tilde{F}_{ab}$ the Hodge-dual of $F_{ab}$. From $L_{\rm boson}$ we
read off the equation of motion for the auxiliary field:
\bea
\frac{\cD}{
\sqrt{1 + g^2 ({\bm u} + {\bar {\bm u}} )+{1 \over 4}g^4 ({\bm u} - {\bar {\bm u}} )^2 } 
} = -\x~.
\eea
Its solution is
\bea
\cD = -\frac{\x}{
\sqrt{1+ g^2 \x^2}
} \,
\sqrt{1 + g^2 (\o + {\bar \o} )+{1 \over 4}g^4 (\o - {\bar \o} )^2 } ~.
\label{D}
\eea
Here the second factor on the right is essentially the BI Lagrangian \cite{BornI}:
\bea
L_{\rm BI} 
&=&   \frac{1}{g^2} \left\{ 1 - \sqrt{1 + g^2 (\o + \bar \o )
+{1 \over 4}g^4 (\o - \bar \o )^2 } 
\right\}
\non \\&=& 
\frac{1}{g^2} \Big\{
1 - \sqrt{- \det (\eta_{ab} + g F_{ab} )} 
\Big\} ~.
\label{BIL}
\eea

Upon elimination of the auxiliary field,  the bosonic Lagrangian (\ref{L-bosonic}) becomes
\bea
L_{\rm boson}=   \frac{1}{g^2} \left\{ 1- 
\sqrt{1+ g^2 \x^2}\, \sqrt{- \det (\eta_{ab} + g F_{ab} )} \right\}~.
\eea
Modulo an irrelevant constant term and an overall normalization factor,   
this is again the BI Lagrangian.

Looking at the expression for the auxiliary field, eq. (\ref{D}), we see that it acquires 
a non-vanishing expectation value
\bea
\langle \cD \rangle = -\frac{\x}{
\sqrt{1+ g^2 \x^2}
} ~.
\eea
Therefore, the manifestly realized supersymmetry of the theory (\ref{S}) becomes 
spontaneously broken. The corresponding supersymmetry transformation of the photino 
is now 
\bea
\d \l_\a ={\rm i} \big( \e Q + {\bar \e} {\bar Q} \big) W_\a\big|_{\q=0} 
= \langle \cD \rangle \,\e_\a ~+~\mbox{field-dependent terms}
\label{Goldstino}
\eea
and therefore the photino turns into a Goldstino.
As a result, the situation is now the following. The model under consideration, eq. (\ref{S}), 
possesses two supersymmetries, of which one ($Q$) is linearly realized, and the other ($S$) 
is nonlinearly realized, as described by eq. (\ref{2nd-var}). On the mass shell, 
both $Q$- and $S$-supersymmetries become  nonlinearly realized. 
Clearly, this does not mean that the $\cN=2$ supersymmetry is completely broken, 
for there is only one Goldstino in the theory. 
Therefore, a special combination of the $Q$- and $S$-supersymmetries must remain unbroken. 
This can be seen explicitly as follows.
The $Q$- and $S$-supersymmetry transformations 
of the theory (\ref{S}) form the $\cN=2$ super-Poincar\'e  algebra without central charge,  
\begin{subequations}
\bea
\{ Q_\a , {\bar Q}_\bd \} &=& 2 (\s^c)_{\a \bd} P_c~, 
\qquad \{ S_\a , {\bar S}_\bd \} = 2 (\s^c)_{\a \bd} P_c~, \\
\{ Q_\a , S_\b \} &=& 0~, \qquad  \qquad \quad  ~\,\{ Q_\a , {\bar S}_\bd \} = 0~,
\eea
\end{subequations}
see \cite{BG,RT} for more detail. 
Now, using the transformation laws (\ref{2nd-var}) and (\ref{Goldstino}), one can readily 
check that $\cN=1$ supersymmetry transformations generated by 
\bea
\cQ_\a := \cos \vf  \, Q_\a  +\sin \vf  \,S_\a ~, \qquad 
\tan \vf = \frac{\x}{
\sqrt{1+ g^2 \x^2} }
\eea
remain unbroken. We see that the result of adding the FI term to the $\cN=1$ supersymmetric 
BI action amounts to a U(1) rotation of the unbroken $\cN=1$ supersymmetry generators in 
the $\cN=2$ super-Poincar\'e  algebra. 
Therefore, the dynamical system (\ref{S}) is a (deformed) vector Goldstone multiplet model for 
partial supersymmetry breaking $\cN=2 \to \cN=1$.
Similarly to the considerations  in \cite{RT,ADM},  the action  (\ref{S}) can be obtained  by 
(i) starting from a non-renormalized  model for an Abelian $\cN=2$ vector multiplet 
with two types (electric and magnetic) of $\cN=2$ FI terms \cite{APT,IZ}, and then (ii) 
integrating out a massive $\cN=1$ scalar multiplet.\footnote{It would be interesting to understand
how to integrate out massive degrees of freedom in the non-Abelian extensions \cite{non-Abel}
of the Antoniadis-Partouche-Taylor model \cite{APT}.}

The supersymmetric BI theory (\ref{sbi}) is known to be invariant under U(1) 
duality rotations \cite{BMZ,KT}. It is in fact a special member of the family 
of models for self-dual nonlinear $\cN=1$ supersymmetric electrodynamics constructed in \cite{KT}
and described by actions of the form: 
\be
S [W]  = \frac{1}{4}\int {\rm d}^6z \, W^2 +
\frac{1}{4}\int {\rm d}^6{\bar z} \,{\bar  W}^2
+  \frac{1}{4}\, \int {\rm d}^8z \, W^2\,{\bar W}^2  \,
\L \big( u , \bar u \big)~, \qquad 
u := \frac{1}{8} D^2 \, W^2~,~~~
\label{gendualaction}
\ee
where $\L(u,\bar u )$ is a real analytic function
of one complex  variable. In accordance with \cite{KT}, this  theory possesses U(1) 
duality invariance provided $\L$ obeys the self-duality equation
\be
{\rm Im}\;  \Big\{ 
\G - \bar u \, \G^2
\Big\} = 0~, 
\qquad \G (u , \bar u ) :=  \frac{\pa  }{\pa u} \Big( u\, \L (u , \bar u )\Big)~.
\label{dif}
\ee
The fermionic sector of such models turns out to possess a remarkable structure  \cite{KMcC}.

As demonstrated in \cite{KMcC}, under the only additional restriction
\be
\L_{u{\bar u}}(0,0) =  3 \L^{3}(0,0) ~, 
\label{cond-for-AV}
\ee
the component fermionic action coincides, modulo a nonlinear field redefinition, 
with the Volkov-Akulov action \cite{VA}.\footnote{The results in   \cite{KMcC} extended
the earlier component analysis  \cite{HK} of  
the supersymmetric BI theory (\ref{sbi}).} 
At first sight, this ubiquity  of the Goldstino action in the framework of nonlinear self-duality
looks somewhat miraculous.
It can be explained, however, if we let the FI term enter the game
and  consider the following model
\be
S [W]  + S_{\rm FI}~.
\label{FI3}
\ee
In the purely bosonic sector, the equation of motion for the auxiliary field is
\bea
\cD \Big[1- \bar {\bm u} \G ({\bm u} , \bar {\bm u}) -  {\bm u}\bar \G  ({\bm u} , \bar {\bm u}) \Big]
= -\x~, 
\label{D2}
\eea 
with $\bm u$ defined as in  (\ref{bm-u}).
This equation should be used to express $\cD$ in terms of the electromagnetic field 
strength,  $\cD = f (\o, \bar \o) $.
Generically, the auxiliary field develops a non-vanishing expectation value, 
$\langle \cD \rangle \neq 0$, which must satisfy an algebraic nonlinear equation that 
follows from (\ref{D2}) by setting $\o=0$.\footnote{In some cases, the algebraic nonlinear equation
on $\langle \cD \rangle $ has no solution, and then (\ref{FI3}) is inconsistent.}   
As a result, the supersymmetry becomes spontaneously broken, 
and thus the photino action should be related to the Goldstino action, 
due to the unique of the latter.

It is worth briefly recalling the structure of the bosonic sector in the model (\ref{gendualaction}).
The corresponding equation of motion for $\cD$ is obtained from (\ref{D2}) by setting $\x=0$, 
and hence it always has the solution $\cD=0$. With this solution chosen, 
the dynamics of the electromagnetic field is described by the Lagrangian
\bea
L = -\hf (\o + \bar \o ) + \o \bar \o \, \L( \o, \bar \o )~, 
\eea
with the interaction $\L$ obeying the self-duality equation (\ref{dif}). 
This is a model for self-dual nonlinear electrodynamics in the sense of \cite{GZ,GR1,GZ2}, 
of which the BI theory (\ref{BIL}) is a special case. Such theories and their generalizations 
possess very interesting properties, see, e.g., the second reference in \cite{KT} 
and \cite{AFZ} for  reviews. 
It is natural to ask the following question: Is self-duality  preserved in some form 
in the case of deformed theory (\ref{FI3}) with  
a non-zero $\cD$ obeying the equation (\ref{D2})?
We now turn to answering this question.

The model (\ref{gendualaction}) can be generalized to include 
couplings to supermultiplets containing the dilaton and axion, 
the NS and RR two-forms, $B_2$ and $C_2$,
and the RR four-form, $C_4$, as presented in  \cite{KT} building on the bosonic constructions 
of \cite{GR2,D-branes,KO}.
The extended action  has the form 
\be
S[W, \F, \b, \g, \O] ~=~ S[\mathbb W , \F]
~+~\left\{  \int {\rm d}^6z \,
\Big(\O +\hf\, \g^\a {\mathbb W}_\a \Big)~+~{\rm c.c.} \right\}~,
\label{super-NS-RR}
\ee
where
\bea
S[W,\F] &=&
\frac{\rm i}{4}\int {\rm d}^6z \, \F\,W^2 -
\frac{\rm i}{4}\int {\rm d}^6{\bar z} \,
{\bar \F}\,{\bar  W}^2  \label{super-d-a}\\
&-&  \frac{1}{16}\, \int {\rm d}^8z \,
(\F-{\bar \F})^2\, W^2\,{\bar W}^2  \,
\L \Big( \frac{\rm i}{16} (\F- \bar \F )\,D^2\,W^2
\, ,\, \frac{\rm i}{16}(\F- \bar \F )\,
{\bar D}^2\, {\bar W}^2 \Big)~~~~ \non
\eea
describes SL$(2,{\mathbb R})$ duality invariant coupling of 
the vector multiplet to the dilaton-axion chiral multiplet $\F$, 
and 
\be
{\mathbb W}_\a :=W_\a + {\rm i}\,\b_\a
\ee
is the supersymmetrization of $F+B$.
Here $\b_\a$, $\g_\a$ and $\O$ are unconstrained chiral superfields
which include, among their components, the fields $B_2$, $C_2$
and $C_4$, respectively. The resulting action is invariant under the following
gauge transformations:
\begin{subequations}
\bea
&\d \b_\a = {\rm i}\, \d W_\a = {\rm i}\,{\bar D}^2 D_\a K_1~, \label{db}\\
&\d \g_\a = {\rm i}\,{\bar D}^2 D_\a K_2,
\qquad
\d \O = \frac{1}{2}\,\b^\a {\bar D}^2 D_\a K_2~,
\\
&\d \O = {\rm i}\, {\bar D}^2 K_3~, \label{6-o}
\eea
\end{subequations}
with $K_i$ real unconstrained superfields.
The action (\ref{super-NS-RR}) reduces to (\ref{gendualaction})
by setting $\F = -{\rm i} $ and switching off the other chiral superfields 
 $\b_\a$, $\g_\a$ and $\O$.

As demonstrated in \cite{KT}, the theory (\ref{super-NS-RR}) is 
invariant under  SL$(2,{\mathbb R})$ 
duality transformations 
\bea
  \left( \begin{array}{c}  M'_\a  \\ W'_\a  \end{array} \right)
~=~  
 \left( \begin{array}{cc} a~& ~b \\ c~ & ~d \end{array} \right) 
\left( \begin{array}{c}  M_\a  \\ W_\a  \end{array} \right) ~,\qquad 
\F' = \frac{a\F +b}{c\F+d}~, \qquad
\left( \begin{array}{cc} a~& ~b \\ c~ & ~d \end{array} \right) \in
{\rm SL}(2, {\mathbb R})~,~~~
\label{N=1dualrot}
\eea
provided the
superfields $\b_\a$, $\g_\a$ and $\O$ transform as
\begin{subequations}
\bea
\left( \begin{array}{c} \g'  \\  \b'  \end{array} \right)
&=&  \left( \begin{array}{cc} a~& ~b \\ c~ & ~d \end{array} \right) 
\left( \begin{array}{c} \g \\ \b  \end{array} \right) ~,
\\
\O' &=& \O -\frac{\rm i}{4}bd \, \b^2
-\frac{\rm i}{2} bc\, \b \g
-\frac{\rm i}{4}ac \,\g^2
~.
\label{Omega}
\eea
\end{subequations}
Here $M_\a$ denotes a variational derivative of the action  with respect to the field strength, 
\be
{\rm i}\,M_\a :=2\, \frac{\d }{\d W^\a}\,S[W, \F, \b, \g, \O] ~.
\ee
A detailed discussion can be found in  \cite{KT}.

Using the second form of the FI term, eq. (\ref{FI-twoforms}), 
the action (\ref{FI3}) is seen to be of the type (\ref{super-NS-RR}) 
with the following ``frozen'' values for background fields: $\g_\a = \x \,\q_\a $, 
$\F = -{\rm i} $  and $\b_\a = \O =0$.
As a natural generalization, an ansatz 
compatible with duality transformations is
$\g_\a \propto \q_\a $,   $\b_\a \propto \q_\a $ and $\O \propto \q^2 $.
A consistent with duality choice is 
\be
\g_\a = \x\, \q_\a ~, \qquad
\b_\a = \z \, \q_\a~, \qquad   \O =0~, \qquad \x, \z \in {\mathbb R}~.
\ee
As follows from (\ref{Omega}),  applying 
a duality transformation generates a purely imaginary non-zero value for
$\O$  which, however,   does not contribute to the action. 

Let us now return to the  model (\ref{FI3}) 
and consider its duality-covariant extension
\be
S [\cW]  + S_{\rm FI}~, \qquad {\cal W}_\a :=W_\a + {\rm i}\,\z\, \q_\a~.
\label{FI4}
\ee
Here the deformed field strength $\cW_\a$ obeys the modified Bianchi identity 
\bea
{\bar D}_\ad {\bar \cW}^\ad - D^\a \cW_\a = 4{\rm i} \z~.
\eea
This action is invariant under inhomogeneous  supersymmetry transformations
\bea
\d W_\a = -{\rm i} \z \e_\a + 
{\rm i} \big( \e Q + {\bar \e} {\bar Q} \big) W_\a~.
\eea
If $S[W]$ coincides with the supersymmetric BI action  (\ref{sbi}), then the resulting model
\bea 
S_{\rm SBI} [\cW] + S_{\rm FI}
\label{sbi3}
\eea
is also invariant under a second nonlinearly realized supersymmetry, 
which is a natural generalization of  (\ref{2nd-var}),
\bea
\d W_\a &=& \eta_\a
+ \frac{g^2}{4} \Big( 
 \frac{1}{4}\, {\bar D}^2 {\bar \cX}\, \eta_\a
+ {\rm i}\,  \pa_{\a \ad} \cX  \, {\bar \eta}^\ad \Big)~,
\eea
where $\cX$ is a chiral superfield, ${\bar D}_\ad \cX=0$,  satisfying
the nonlinear constraint
\be
\cX +\frac{g^2}{16} \,  \cX\, {\bar D}^2  
{\bar \cX}  = \cW^2~,
\ee
compare with (\ref{n=1constraint}). Our action (\ref{sbi3}) is a two-parameter deformation 
of the supersymmetric BI theory (\ref{sbi}). This action appears to be the most general 
Goldstone-Maxwell multiplet model for partial $\cN=2 \to \cN=1$
supersymmetry breaking. Requiring the first supersymmetry to be manifest
eliminates one of the deformations, $\z=0$. The supersymmetric BI action (\ref{sbi}) is indeed unique
if the first supersymmetry is required to be  unbroken.

In conclusion, we would like to comment on the statement  made in \cite{DT} that 
``no supercurrent supermultiplet exists for globally supersymmetric gauge theories with non-zero 
Fayet-Iliopoulos terms.'' A general scheme to compute supercurrents \cite{FZ}
in rigid supersymmetric theories is by evaluating a variational derivative of the action functional 
with respect to the gravitational superfield $H^{\a \ad}$, 
\be
J_{\a \ad} = \frac{\D S}{\D H^{\a \ad}}~,
\label{supercurrent}
\ee
with the idea due to Ogievetsky and Sokatchev \cite{OS}.
More specifically, the procedure is as follows: (i) one should lift the theory to a curved 
superspace corresponding to one of the known off-shell supergravity formulations 
(realized in terms of  the gravitational superfield $H^{\a \ad}$ and an appropriate  compensator); 
(ii) compute the (covariantized)  variational derivative $\D S/\D H^{\a \ad}$; 
(iii) return to the flat  superspace by switching off the supergravity prepotentials.
The scheme is worked out in detail in two textbooks \cite{BK,GGRS}, including numerous examples.
The explicit form of the supercurrent conservation equation depends on the off-shell supergravity 
realization chosen. 
In the case of the old minimal ($n=-1/3$) formulation  for $\cN=1$ supergravity \cite{old}
(see also \cite{SG}),
the conservation equation
is 
\be
{\bar D}^{\ad}J_{\a \ad} = D_\a T~,\qquad {\bar D}_\ad T =0~,
\label{conservation-old}
\ee
with $T$ called the supertrace. 
 In the case of the new minimal ($n=0$) formulation  for $\cN=1$ supergravity \cite{new}, 
the conservation equation is 
\be
{\bar D}^{\ad}{\mathbb J}_{\a \ad} = {\mathbb T}_\a ~,
\qquad {\bar D}_\ad {\mathbb T}_\a  =0~, 
\qquad D^\a {\mathbb T}_\a = {\bar D}_\ad {\bar {\mathbb T}}^\ad~.
\label{conservation-new}
\ee
The difference between (\ref{conservation-old}) and (\ref{conservation-new}) is due to 
the different types of compensators used in these supergravity formulations. 
One can construct more general supercurrent conservation equations, 
see, e.g., \cite{PS,CL}.

We wish to analyze the two versions of the supercurrent for the theory (\ref{FI3}).  
Within the old minimal formulation for $\cN=1$ supergravity, the supercurrent for the model
(\ref{gendualaction}) was computed in  \cite{KMcC1} and shown to be {\it duality invariant}. 
The supercurrent obtained is rather complicated  to deal with. So for simplicity, 
we restrict our consideration to the Maxwell case, by setting $\L =0$ in (\ref{gendualaction}), 
similarly to \cite{KS}. The consideration below can naturally be generalized to the case $\L\neq 0$.

Using  the old minimal formulation for $\cN=1$ supergravity, 
the supermultiplets $J_{\a \ad } $ and $T$ are computed to be \cite{KS}:
\bea
J_{\a \ad} = 2 W_\a {\bar W}_\ad +\frac{2}{3} \x [D_\a, {\bar D}_\ad ] V~, 
\qquad 
T = \frac{1}{3} \x {\bar D}^2 V~.
\label{JT1} 
\eea 
Using the equation of motion, $D^\a W_\a = 2 \x$, one can check 
that the conservation equation (\ref{conservation-old}) holds. 
As pointed out by Komargodski and Seiberg \cite{KS}, both 
$J_{\a \ad } $ and $T$ are not gauge invariant. The reason for this 
is very simple: minimal coupling of the FI term to supergravity, 
which makes use of the first expression in  (\ref{FI-twoforms}), 
is not gauge invariant.  Old minimal supergravity is not well suited to describe  FI terms.

Using  the new minimal formulation for $\cN=1$ supergravity, 
the supermultiplets ${\mathbb J}_{\a \ad } $ and ${\mathbb T}_\a$ can be computed to be:
\bea
{\mathbb J}_{\a \ad} = 2 W_\a {\bar W}_\ad ~, 
\qquad 
{\mathbb T} _\a = 4 \x W_\a~.
\label{JT2} 
\eea 
Both objects are gauge invariant. The reason for this 
is very simple: minimal coupling of the FI term to supergravity, 
which makes use of the second expression in  (\ref{FI-twoforms})
and is obtained by replacing (the vacuum expectation value) 
$ \q^\a$ with the chiral spinor compensator, 
is gauge invariant. New minimal supergravity is ideal for describing  FI terms.

The theory (\ref{FI3}) is $R$-invariant, and the corresponding $R$-current $j^{(5)}_{\a \ad}$
must be conserved. 
If one identifies $j^{(5)}_{\a \ad}$ with the lowest component of the supercurrent in
(\ref{JT1}), it fails to be conserved, for one finds 
\bea
\pa^{\a \ad} J_{\a \ad} 
= \frac{\rm i}{6} \x \,  [D^2 ,  {\bar D}^2 ] V 
\equiv \frac{2}{3} \x \, \pa^{\a \ad} [D_\a ,  {\bar D}_\ad ] V \neq 0~. 
\eea
This result was interpreted in \cite{DT} as non-existence of
 supercurrent  for rigid supersymmetric gauge theories 
 with non-zero FI terms. 
However, if one defines the  $R$-current  by
\bea
j^{(5)}_{\a \ad} :=  \Big( J_{\a \ad}  - \frac{2}{3} \x \,  \big[ D_\a ,  {\bar D}_\ad \big]  V \Big) 
\Big|_{\q =0}~, 
\eea
then it is clearly conserved. 

In the case of the supercurrent (\ref{JT2}),
the standard definition of the  $R$-current applies 
\bea
j^{(5)}_{\a \ad} :=  {\mathbb J}_{\a \ad}|_{\q =0}~.
\eea  
Our consideration shows that the supercurrent 
does exist for globally supersymmetric gauge theories 
in the presence of FI  terms.\\

\noindent
{\bf Acknowledgements:}\\
The author is grateful to Ian McArthur and Arkady Tseytlin for useful comments.
This work is supported  in part by the Australian Research Council.\\

\noindent
{\bf Note added.} The published version of \cite{DT}, Ref. \cite{DT2}, significantly differs
from the original preprint, Ref.  \cite{DT}, discussed above. The conclusions
of \cite{DT2} are similar to those given  in the present paper.

\small{

}

\end{document}